\documentclass[conference,letterpaper]{IEEEtran}

\usepackage{amsmath,amssymb}
\usepackage{booktabs}
\usepackage{graphicx}
\usepackage{xcolor}
\colorlet{revisioncolor}{black}
\newcommand{\revision}[1]{\textcolor{revisioncolor}{#1}}
\colorlet{newrevisioncolor}{black}
\newcommand{\newrev}[1]{\textcolor{newrevisioncolor}{#1}}
\colorlet{currentrevisioncolor}{black}
\newcommand{\octrev}[1]{\textcolor{currentrevisioncolor}{#1}}
\usepackage{algorithm}
\usepackage{algorithmic}
\usepackage{cite}
\usepackage{microtype}
\usepackage[hidelinks,bookmarks=false]{hyperref}

\makeatletter
\renewcommand\fs@ruled{\def\@fs@cfont{\bfseries}\let\@fs@capt\floatc@ruled
  \def\@fs@pre{\kern3pt\hrule height.8pt depth0pt \kern1pt}%
  \def\@fs@post{\hrule\relax}%
  \def\@fs@mid{\kern1.5pt\hrule\kern1.5pt}%
  \let\@fs@iftopcapt\iftrue}
\makeatother

\newcommand{\AI}{\mathrm{A}}
\newcommand{\DD}{\mathrm{D}}
\newcommand{\setA}{\mathcal{K}_{\AI}}
\newcommand{\schD}{\mathcal{K}_{\DD}}
\newcommand{\resultpanelwidth}{0.239\textwidth}

\title{Analog RF Computing for Integrated Data and AI Flows Over Wireless Networks}

\author{\IEEEauthorblockN{Wentao Yu and Vincent W.S. Wong}
\IEEEauthorblockA{Department of Electrical and Computer Engineering, The University of British Columbia, Vancouver, BC, Canada\\
Email: \{wentaoyu, vincentw\}@ece.ubc.ca}}

\begin{document}
\maketitle
\linespread{0.97}\selectfont
\tolerance=1000
\emergencystretch=0.5em
\setlength{\abovedisplayskip}{1.2pt plus 1pt minus 0.6pt}
\setlength{\belowdisplayskip}{1.2pt plus 1pt minus 0.6pt}
\setlength{\abovedisplayshortskip}{0.8pt plus 0.8pt minus 0.4pt}
\setlength{\belowdisplayshortskip}{0.8pt plus 0.8pt minus 0.4pt}

\begin{abstract}
Many resource-constrained edge devices require artificial intelligence (AI) services but are not powerful enough to store or execute neural networks (NNs). Analog radio frequency (RF) computing addresses this limitation by letting a base station (BS) multicast a waveform encoding NN weights to different clients. Each client then multiplies the received weight-encoded waveform with its locally generated input-encoded waveform through \octrev{a} passive mixer in its RF frontend, thereby realizing the matrix-vector multiplications (MVMs) needed for AI inference with an extremely low power and memory requirement. This can enable telecom operators to provide AI inference services to the edge devices connected to their networks. In practice, however, this multicast AI flow must share the spectrum, power, and spatial resources with other data flows. This paper considers a multi-user multiple-input multiple-output (MU-MIMO) downlink \octrev{scenario} and studies the resulting relative aggregate data rate reduction from integrating the multicast AI flow with the unicast data flows. We formulate a joint beamforming problem for the data and AI flows to maximize the aggregate data rate subject to computing accuracy, client energy, total power, and hardware-related constraints. We develop a weighted minimum mean-square error (WMMSE)-based algorithm to solve the problem. Numerical results based on \octrev{3rd Generation Partnership Project (}3GPP\octrev{)} channels show that the relative aggregate data rate reduction from integrating an AI flow with data flows is below 5\% in representative settings. This indicates that telecom operators can use analog RF computing to provide edge AI services while providing data communication services simultaneously. 
\end{abstract}

\section{Introduction}
Wireless networks are traditionally designed to provide data communication services. As telecommunications markets mature, some telecom operators are \octrev{launching} artificial intelligence (AI) services as a new source of revenue. AI flow \cite{shao2025ai,an2026ai} is a framework for ubiquitous edge intelligence that integrates generative transmission and familial models to support efficient device-edge-cloud collaboration under limited edge computing and bandwidth resources. It provides a path for telecom operators to deliver AI capabilities alongside data communication services.

In this paper, we focus on how telecom operators can enable AI inference for resource-constrained edge devices such as Internet of things (IoT) \octrev{devices}, cameras, wearables, and sensors. Digital on-device inference requires storing the neural network (NN) model and executing it on digital processors, such as graphics processing units (GPUs). The memory and energy requirements for AI inference may exceed the capabilities of some edge devices \cite{sze2017efficient}. To this end, offloading an AI inference task to an edge or cloud provider can mitigate on-device computation. However, it increases the communication overhead and may expose private information. The edge or cloud provider's digital hardware performs the inference, while the telecom operator still only serves as a data pipeline and cannot capture revenue from computing. We therefore ask whether the wireless network itself can support AI inference, allowing telecom operators to offer AI inference services directly to resource-constrained devices. 

\begin{figure}[t]
\centering
\includegraphics[hiresbb,width=0.98\columnwidth]{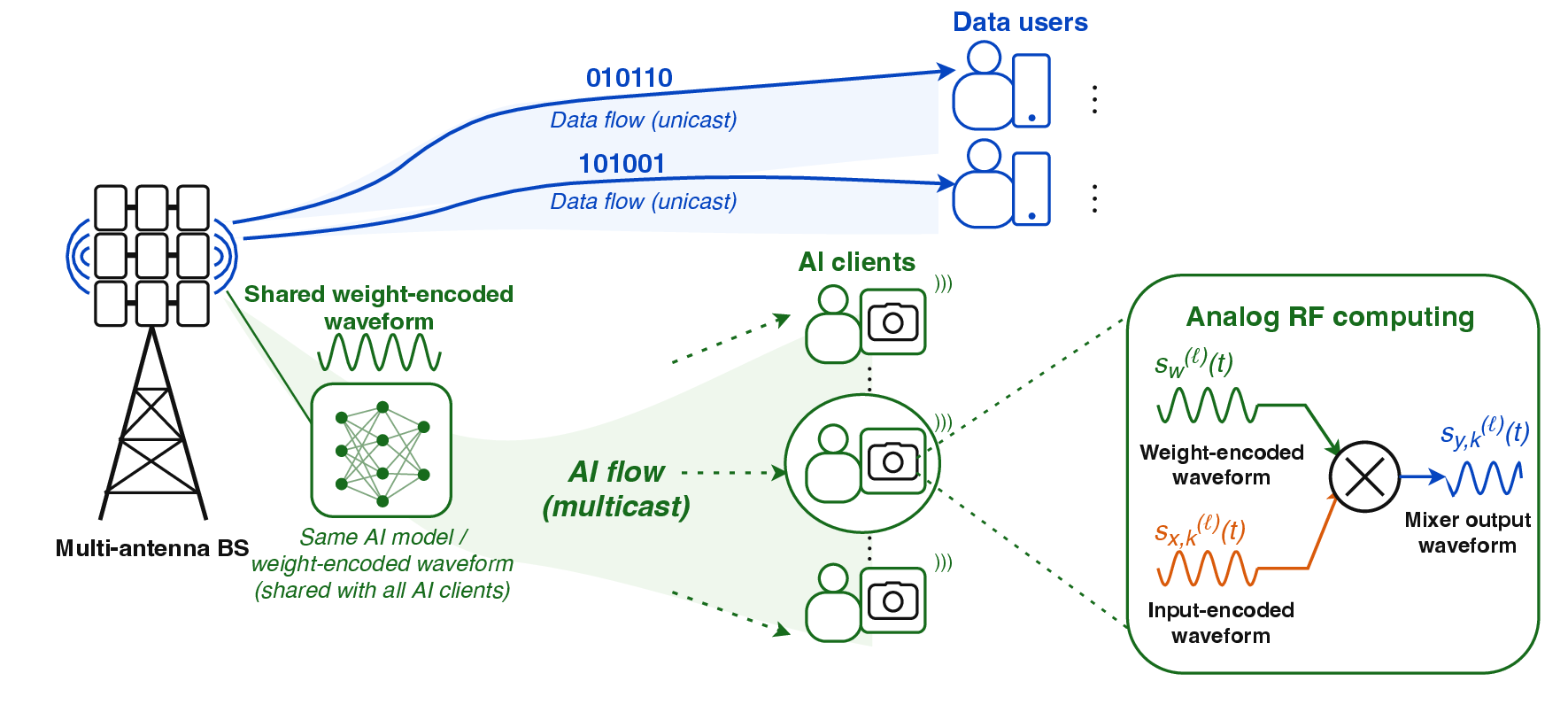}
\caption{A multi-antenna BS serves data users \octrev{via} unicast data flows and AI clients \octrev{via} a multicast weight\octrev{-encoded} waveform for analog RF computing.}
\label{fig:system}
\end{figure}

\begin{figure*}[t]
\centering
\includegraphics[hiresbb,width=0.90\textwidth]{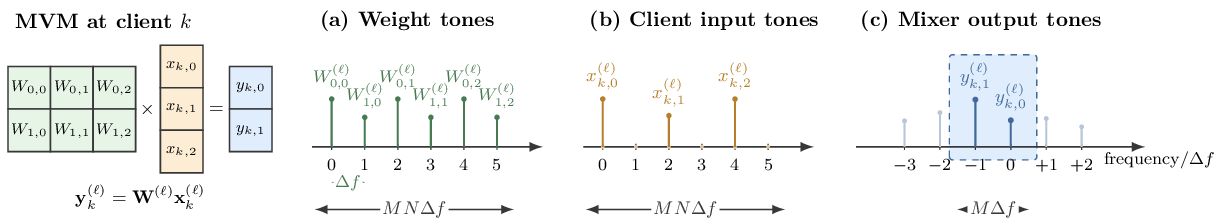}
\caption{Example for mixer-based MVM with $M=2$ and $N=3$. (a) Weight $W_{m,n}^{(\ell)}$ modulates tone $nM+m$. (b) Input $x_{k,n}^{(\ell)}$ modulates tone $nM$. (c) Difference-frequency products for row $m$ align at $-m\Delta f$ and form $y_{k,m}^{(\ell)}$. }
\label{fig:mvm_mapping}
\end{figure*}

Radio frequency (RF) signals in wireless systems have been shown to enable energy-efficient analog computing. Existing methods utilize the superposition property of RF signals for over-the-air aggregation or use reconfigurable microwave networks for predefined linear functions \cite{goldenbaum2013robust,sanchez2023airnn,wen2024taskoriented,reusmuns2023airfc,nerini2025analog}. They are often confined to specific linear operations and hardware configurations. On the other hand, \emph{analog RF computing} enables RF signals to perform general matrix-vector multiplications (MVMs) \cite{yu2026analogrf,gao2026wise}. Thus, the wireless communication network can provide AI inference services. 

In analog RF computing, the base station (BS) broadcasts a multicarrier waveform encoded with an NN weight matrix. Each AI client uses a passive mixer in its RF frontend to multiply its input-encoded waveform with the received weight-encoded waveform. The MVM output is recovered after filtering and readout. Repeating this process across NN layers enables on-device AI inference with ultra-low energy and memory consumptions. Multiple AI clients can receive the same multicast weight-encoded waveform and multiply it with its own inputs to obtain client-specific inference results. 

However, wireless systems must continue to provide data communication services. At the BS, one multicast AI stream\footnote{This paper considers one AI model at the BS. The AI flow is a single-group multicast service. Supporting multiple AI models would lead to multi-group multicast beamforming and is left for future work.}, i.e., the physical-layer signal of the AI flow, carries the weight-encoded waveform, while several unicast data streams carry user data. These streams are transmitted over shared wireless resources. This will create interference. The AI stream interferes with data users, while the data streams perturb the MVM results at AI clients. Thus, the beamformers for the AI and the data streams must be jointly designed. In Fig.~\ref{fig:system}, we illustrate the wireless system for integrated data and AI flows. 

In our previous work \cite{yu2026analogrf}, we proposed analog RF computing-based edge AI over multi-user multiple-input multiple-output (MU-MIMO) systems and considered a computing-centric physical layer in which the BS broadcasts only an AI stream without concurrent data streams. In this paper, we consider a more practical wireless system by providing integrated data and AI services. We aim to support one multicast AI stream and multiple data streams without increasing the transmit power budget. We quantify its relative aggregate data rate reduction and examine the effects of key system parameters. The main contributions are as follows.
\begin{itemize}
\setlength{\itemsep}{1pt}
\setlength{\parskip}{0pt}
\setlength{\parsep}{0pt}
\setlength{\topsep}{2pt}
\item We model integrated data and AI flows in which one multicast AI stream and multiple unicast data streams share \octrev{the} wireless resources. \octrev{We compare the integrated system with a} data-flow-only reference \octrev{under} the same data users, channels, bandwidth, and BS power \octrev{to quantify} the aggregate data rate reduction \octrev{due to} the AI flow.
\item We formulate a sum-rate maximization problem subject to MVM computing accuracy, client energy, BS transmit power, and hardware constraints of mixers. We develop a weighted minimum mean-square error (WMMSE) algorithm that jointly designs the AI and data beamformers. 
\item \looseness=-1 Under \octrev{3rd Generation Partnership Project (}3GPP\octrev{)} channels, we demonstrate how the computing accuracy target, spatial degrees of freedom, number of AI clients, and client energy limit determine the relative aggregate data rate reduction of integrated data and AI flows. Under representative system settings, integrated data and AI flows only lead to a relative aggregate data rate reduction of less than 5\% compared to the data-flow-only reference, which indicates that telecom operators can provide AI inference services while largely preserving the quality of data communication\octrev{s}. 
\end{itemize}

\section{Analog RF Computing for AI Flow}
\subsection{Mixer-Based MVM}
NN inference consists of repeated linear operations followed by lightweight nonlinear activation functions. The linear operations account for most of the computation and can be represented as MVMs. Analog RF computing performs each MVM in the receiver RF frontend, after which the nonlinear activation function is applied to the readout. In an AI flow, the BS broadcasts a weight-encoded waveform, while each client generates an input-encoded waveform. The mixer multiplies these two time-domain waveforms. Fig.~\ref{fig:mvm_mapping} depicts their frequency-domain tones. Since multiplication in time corresponds to convolution in frequency, each tone pair produces sum- and difference-frequency components. The tones are carefully assigned such that the products for each output row align at a different difference frequency. Afterwards, filtering will remove the other unwanted products before readout.

For NN layer $\ell\in\mathcal{L}\triangleq\{1,\ldots,L\}$ and AI client $k\in\setA\triangleq\{1,\ldots,K_{\AI}\}$, let
$\mathbf{W}^{(\ell)}\in\mathbb{C}^{M^{(\ell)}\times N^{(\ell)}}$
denote the weight matrix and $\mathbf{x}_{k}^{(\ell)}\in\mathbb{C}^{N^{(\ell)}}$ denote the input.
The output of AI client $k$ at NN layer $\ell$, denoted by $\mathbf{y}_{k}^{(\ell)}\in\mathbb{C}^{M^{(\ell)}}$, is
\begin{equation}
\mathbf{y}_{k}^{(\ell)}
=\mathbf{W}^{(\ell)}\mathbf{x}_{k}^{(\ell)}.
\label{eq:mvm}
\end{equation}
\octrev{Its $m$-th entry is denoted by $y_{k,m}^{(\ell)}$.}
Let $M=M^{(\ell)}$, $N=N^{(\ell)}$, where $m\in\{0,\ldots,M-1\}$ is the row index and $n\in\{0,\ldots,N-1\}$ is the column index, and let adjacent tones be separated by $\Delta f$. The frequency mapping in \cite{yu2026analogrf} forms
\begin{align}
s_w^{(\ell)}(t)
&=\sum_{n=0}^{N-1}\sum_{m=0}^{M-1}
W_{m,n}^{(\ell)}
e^{\mathrm{j}2\pi(nM+m)\Delta f t},
\label{eq:weight_wave}\\
s_{x,k}^{(\ell)}(t)
&=\sum_{n=0}^{N-1}x_{k,n}^{(\ell)}
e^{\mathrm{j}2\pi nM\Delta f t},
\quad 0\le t<1/\Delta f .
\label{eq:input_wave}
\end{align}
Specifically, $W_{m,n}^{(\ell)}$ occupies tone $nM+m$, whereas $x_{k,n}^{(\ell)}$ occupies tone $nM$. Selecting the input-tone frequency minus the weight-tone frequency places their product at $-m\Delta f$, independent of column index $n$. All products from row $m$ of $\mathbf{W}^{(\ell)}$ consequently add at the same output tone, where $\propto$ denotes proportionality:
\begin{equation}
s_{y,k}^{(\ell)}(t)\propto
\sum_{m=0}^{M-1}
\left(\sum_{n=0}^{N-1}W_{m,n}^{(\ell)}x_{k,n}^{(\ell)}\right)
e^{-\mathrm{j}2\pi m\Delta f t}
+s_{\mathrm{other}}(t).
\label{eq:mixer_output}
\end{equation}
A bandpass filter rejects $s_{\mathrm{other}}(t)$, and a discrete Fourier transform (DFT) readout extracts the $M$ selected tones as $\mathbf{y}_{k}^{(\ell)}$. The dark output tones in Fig.~\ref{fig:mvm_mapping} are the selected components. Products across the columns of each row align at one output tone. If a layer has more output rows than the hardware can support simultaneously, then the rows are processed in blocks. The BS multicasts one weight-encoded waveform, while each client supplies its input-encoded waveform.

The BS multicasts the NN weight matrices layer by layer. The physical layer design is fixed within each layer but is only reconfigured between layers. We therefore formulate the design for one layer and omit the index $\ell$ \octrev{below}.

For a BS with $N_{\mathrm t}$ transmit antennas, the MVM amplitude at AI client $k$ contains a BS-controlled factor and a client-controlled factor. Let $\mathbf{h}_{\AI,k}\in\mathbb{C}^{N_{\mathrm t}}$ denote the channel between the BS and AI client $k$, let $\mathbf{f}_{\AI}\in\mathbb{C}^{N_{\mathrm t}}$ denote the AI beamformer, and let $P_{w,0}$ denote the reference power of weight-encoded waveform. The received computing gain is
\begin{equation}
g_k=\frac{\mathbf{h}_{\AI,k}^{\mathsf H}\mathbf{f}_{\AI}}{\sqrt{P_{w,0}}}.
\label{eq:gain}
\end{equation}
\newrev{Client $k$ applies $\beta_k\in\mathbb{C}$ to the complex baseband envelope of its input-encoded waveform, making the mixer output proportional to $g_k\beta_k$. The AI beamformer and client scaling determine computing accuracy and client energy consumption.}


\subsection{Computing Accuracy Model}
The receiver noise and interference from the data flows produce independent zero-mean errors, which contribute additively to the \octrev{normalized mean-square error (}NMSE\octrev{)}. For the data-user set $\schD=\{1,\ldots,K_{\DD}\}$, the data-flow interference power at AI client $k$ is
\begin{equation}
I_{\AI,k}(\mathbf{F}_{\DD})
=\frac{1}{P_{w,0}}\sum_{j\in\schD}
|\mathbf{h}_{\AI,k}^{\mathsf H}\mathbf{f}_{\DD,j}|^2,
\label{eq:data_interference}
\end{equation}
where $\mathbf{f}_{\DD,j}\in\mathbb{C}^{N_{\mathrm t}}$ is the data beamformer for user $j\in\schD$, and $\mathbf{F}_{\DD}\triangleq[\mathbf{f}_{\DD,1},\ldots,\mathbf{f}_{\DD,K_{\DD}}]\in\mathbb{C}^{N_{\mathrm t}\times K_{\DD}}$ is the collection of data beamformers. The NMSE of the analog MVM is
\begin{equation}
\zeta_k
=\underbrace{\frac{1}{\kappa|g_k\beta_k|^2}}_{\text{receiver-noise error}}
+\underbrace{\frac{I_{\AI,k}(\mathbf{F}_{\DD})}{|g_k|^2}}_{\text{data-flow interference error}}\revision{\mathpunct{.}}
\label{eq:nmse}
\end{equation}
The \revision{coefficient $\kappa$ in the receiver-noise error} is given by \cite{yu2026analogrf}:
\begin{equation}
\kappa
=\frac{\rho_{\mathrm{mixer}}\rho_{\mathrm{nf}}P_{w,0}P_{x,0}}
{k_{\mathrm B}T_0B},
\label{eq:kappa}
\end{equation}
where $P_{x,0}$ is the reference power of \octrev{the} input-encoded waveform, $B$ is \octrev{the} bandwidth, $k_{\mathrm B}$ is \octrev{the} Boltzmann constant, $T_0$ is the receiver noise temperature, $\rho_{\mathrm{mixer}}$ is the mixer conversion coefficient, and $\rho_{\mathrm{nf}}$ is the receiver-noise coefficient. Increasing $|g_k|$ suppresses both terms of the NMSE, while increasing $|\beta_k|$ only reduces the error caused by receiver noise. The computing-accuracy requirement is given by $\sqrt{\zeta_k}\le\epsilon_k$, equivalently $\zeta_k\le\epsilon_k^2$, where $\epsilon_k$ is the root-NMSE target.

\subsection{Client Energy Consumption}
The client generates its input-encoded waveform, samples the mixer output, and readouts the MVM result. We \octrev{normalize} its energy consumption \octrev{by the number of} multiply-accumulate (MAC) operation\octrev{s}. Only the input-waveform generation depends on $\beta_k$, while sampling and readout form a fixed per-MAC floor. Following the model in \cite{yu2026analogrf}, the energy consumption \octrev{per MAC} of AI client $k$ is a summation of three terms, which is given by
\begin{equation}
e_{\AI,k}
=e_{1,k}+e_2+e_3.
\label{eq:energy_sum}
\end{equation}
\octrev{The total client energy for one MVM is $MN e_{\AI,k}$.}

\begin{algorithm}[t]
\caption{WMMSE-Based Joint Beamforming}
\label{alg:main}
\begin{algorithmic}[1]
\footnotesize
\setlength{\baselineskip}{8.5pt}
\STATE \textbf{Input:} $\mathbf{H}_{\AI}$, $\mathbf{H}_{\DD}$, $P_{\max}$, $P_{\mathrm{LO,th}}$, $\mathcal V$, and client parameters.
\STATE Compute $\bar\beta$, $\eta$, and $q_{k,\min}$ using \eqref{eq:beta}, \eqref{eq:computing}, and \eqref{eq:budget}.
\FOR{$v\in\mathcal V$}
\STATE Set $q_k(v)$ using \eqref{eq:power_grid} and compute \eqref{eq:aiclosed}.
\STATE Form data-flow interference budgets \newrev{$b_k(q_k(v))$}.
\STATE Initialize data beamformers by regularized zero forcing.
\REPEAT
\STATE Update $u_j$ and $\mu_j^{\mathrm{mmse}}$ using \eqref{eq:mmse}, and set $\omega_j=(\mu_j^{\mathrm{mmse}})^{-1}$.
\STATE Update data beamformers using \eqref{eq:kkt} and projected dual search.
\STATE Project onto the interference and total-power sets.
\UNTIL{the relative sum-rate change is below \octrev{a} tolerance \octrev{level}}
\ENDFOR
\STATE Return the feasible candidate with largest $R$.
\end{algorithmic}
\end{algorithm}
The controllable waveform-generation term is
\begin{equation}
e_{1,k}
=\xi P_{x,0}|\beta_k|^2,\quad
\xi
=\frac{(1+\vartheta)(1+\varpi)}
{4B\rho_{\mathrm{radio}}},
\label{eq:energy_input}
\end{equation}
where $\vartheta$ is the guard factor, $\varpi$ is the cyclic prefix overhead, and $\rho_{\mathrm{radio}}$ is client RF chain efficiency. With $N$ input columns and $M^{\prime}$ output rows per block, the analog-to-digital converter (ADC) and digital readout terms are
\begin{align}
e_2
&=\frac{1+\vartheta}{2N}e_{\mathrm{adc}},
\label{eq:energy_adc}\\
e_3
&=\frac{1+\vartheta}{2N}
\log_2\!\big((1+\vartheta)M^{\prime}\big)e_{\mathrm{dec}}.
\label{eq:energy_read}
\end{align}
Here $e_{\mathrm{adc}}$ is the energy per ADC sample and $e_{\mathrm{dec}}$ is the energy per digital readout operation. To compare the client energy consumption with digital inference using $e_{\mathrm{digital}}$ as the digital inference energy per MAC and $\tau_E$ as the analog energy target relative to this reference, we require
\begin{equation}
e_{\AI,k}\le\tau_E e_{\mathrm{digital}},\quad
k\in\setA,\qquad 0<\tau_E\le1\mathpunct{.}
\label{eq:energy_target}
\end{equation}
For example, $\tau_E=0.1$ sets the analog energy target to $10\%$ of a digital per-MAC reference. We denote by $\bar\beta\in\mathbb{R}_{+}$ the upper bound on $|\beta_k|$ imposed by \eqref{eq:energy_sum}--\eqref{eq:energy_target} and the RF input limit $|\beta_k|\le\beta_{\max}$\newrev{, where $\beta_{\max}\in\mathbb{R}_{+}$}:
\begin{equation}
\bar{\beta}\triangleq
\min\!\left\{\beta_{\max},
\sqrt{\frac{\tau_E e_{\mathrm{digital}}-e_2-e_3}
{\xi P_{x,0}}}\right\}.
\label{eq:beta}
\end{equation}
Hence, $|\beta_k|\le\bar\beta$ should hold for every $k\in\setA$. A smaller $\tau_E$ limits the client-side energy consumption and shifts more of the computing-gain requirement to the BS.

\section{Integrated Data and AI Flows: System Model and Problem Formulation}
\subsection{System Model}
\octrev{A} fully digital BS with $N_{\mathrm t}$ antennas and $N_{\mathrm{RF}}=N_{\mathrm t}$ RF chains serves $K_{\AI}$ single-antenna AI clients, indexed by $\setA$, through one multicast AI stream and serves $K_{\DD}$ single-antenna data users in $\schD$ through unicast data streams. In the same time-frequency resource, the transmitted signal $\mathbf{x}_{\mathrm{BS}}\in\mathbb{C}^{N_{\mathrm t}}$ is
\begin{subequations}\label{eq:tx}
\begin{gather}
\mathbf{x}_{\mathrm{BS}}
=\mathbf{f}_{\AI}s_w+\sum_{j\in\schD}\mathbf{f}_{\DD,j} s_j,
\label{eq:tx_signal}\\
\|\mathbf{f}_{\AI}\|^2+\sum_{j\in\schD}\|\mathbf{f}_{\DD,j}\|^2
\le P_{\max}.
\label{eq:tx_power}
\end{gather}
\end{subequations}
The number of data users remains $K_{\DD}$ in both the data-flow-only and integrated systems. Let $\mathbf{h}_{\DD,j}\in\mathbb{C}^{N_{\mathrm t}}$ denote the channel between the BS and data user $j$.
Let $n_j\sim\mathcal{CN}(0,\sigma_j^2)$ denote receiver thermal noise, where $\sigma_j^2=k_{\mathrm B}T_0B/\rho_{\mathrm{nf}}$. 
The signal-to-interference-plus-noise ratio (SINR) of data user $j$ is
\begin{equation}
\gamma_j=
\frac{|\mathbf{h}_{\DD,j}^{\mathsf H}\mathbf{f}_{\DD,j}|^2}
{\sum_{i\ne j}|\mathbf{h}_{\DD,j}^{\mathsf H}\mathbf{f}_{\DD,i}|^2+
|\mathbf{h}_{\DD,j}^{\mathsf H}\mathbf{f}_{\AI}|^2+\sigma_j^2}.
\label{eq:sinr}
\end{equation}
The term $|\mathbf{h}_{\DD,j}^{\mathsf H}\mathbf{f}_{\AI}|^2$ is the interference caused by the AI stream at data user $j$. The reverse direction is interference from the data flows to the AI clients, as discussed in \eqref{eq:data_interference}. The objective for communication performance is the aggregate data rate
\begin{equation}
R=B\sum_{j\in\schD}\log_2(1+\gamma_j).
\label{eq:sumrate}
\end{equation}

\subsection{Problem Formulation}
For fixed BS beamformers, increasing $|\beta_k|$ monotonically reduces the noise term in \eqref{eq:nmse}. Hence, $|\beta_k|=\bar\beta$ is optimal and eliminates the client-side variable $\beta_k$. Substituting \eqref{eq:gain} and \eqref{eq:data_interference} into \eqref{eq:nmse} then gives the beamformer-only computing constraint
\begin{equation}
\epsilon_k^2|\mathbf{h}_{\AI,k}^{\mathsf H}\mathbf{f}_{\AI}|^2
-\sum_{j\in\schD}|\mathbf{h}_{\AI,k}^{\mathsf H}\mathbf{f}_{\DD,j}|^2
\ge\eta,\quad
\eta\triangleq\frac{P_{w,0}}{\kappa\newrev{\bar\beta^2}}.
\label{eq:computing}
\end{equation}
The factor $|\mathbf{h}_{\AI,k}^{\mathsf H}\mathbf{f}_{\AI}|^2$ is the received power of the weight-encoded AI waveform and quantifies the BS-controlled AI-flow strength available for mixer-based MVM. We refer to it as the received computing power. Constraint \eqref{eq:computing} states that this power, scaled by the accuracy tolerance, must exceed the aggregate data-flow interference by the threshold $\eta$. \octrev{Also}, as a hardware constraint, the local oscillator (LO) port of \octrev{the} passive mixer must be in its small-signal region \cite{yu2026analogrf}:
\begin{equation}
|\mathbf{h}_{\AI,k}^{\mathsf H}\mathbf{f}_{\AI}|^2
\le P_{\mathrm{LO,th}},\qquad k\in\setA.
\label{eq:lo}
\end{equation}
We maximize the aggregate data rate subject to the BS power constraint, the beamformer-only computing constraint, and the hardware constraint of passive mixers:
\begin{align}
\mathrm{P1:}\quad
\underset{\mathbf{f}_{\AI},\mathbf{F}_{\DD}}{\operatorname{maximize}}\quad&
R
\notag\\
\operatorname{subject\ to}\quad&
\eqref{eq:tx_power},\ \eqref{eq:computing},\ \eqref{eq:lo}.
\label{eq:problem}
\end{align}
Problem P1 is non-convex because the SINRs are coupled and constraint \eqref{eq:computing} contains a difference of quadratic terms. 


Let $R_{\mathrm{data}}$ denote the aggregate data rate for the data-flow-only reference, which is obtained by setting $\mathbf{f}_{\AI}=\mathbf{0}$ and optimizing the \revision{aggregate data rate} in \eqref{eq:sumrate} over the same $K_{\DD}$ data users. Let $R_{\mathrm{int}}$ denote the value of that same expression at the beamformers returned for problem P1. Both rates therefore refer to identical data users, channels, bandwidth, and BS power, while only $R_{\mathrm{int}}$ includes the AI flow and its constraints. The aggregate data rate reduction $\Delta R$ and relative aggregate data rate reduction $\delta_R$ are denoted as 
\begin{equation}
\Delta R=R_{\mathrm{data}}-R_{\mathrm{int}},\qquad
\delta_R=\frac{\Delta R}{R_{\mathrm{data}}}.
\label{eq:cost}
\end{equation}
Here, $\Delta R$ and $\delta_R$ measure the absolute and relative aggregate data rate reductions, respectively, imposed by the AI flow over the data flows of the $K_{\DD}$-user data communication service. 

\section{WMMSE-Based Joint Beamforming}
Problem P1 couples the AI beamformer and data beamformers through the non-convex computing constraint in \eqref{eq:computing}. We handle this coupling by searching over a finite set of target received computing powers. Each target determines a multicast AI beamformer and a set of data-flow interference budgets, after which the data beamformers are updated via WMMSE iterations. The feasible candidate with the largest data rate is retained as the solution to the problem. The overall procedure is summarized in Algorithm 1. 

\subsection{Search of Received Computing Power}
Define the received computing power at AI client $k$ as
$q_k=|\mathbf{h}_{\AI,k}^{\mathsf H}\mathbf{f}_{\AI}|^2$. It is the AI-flow power at the mixer input, while $\beta_k$ controls the input-encoded waveform at the AI client. Constraint \eqref{eq:computing} then becomes
\begin{equation}
\begin{aligned}
\sum_{j\in\schD}|\mathbf{h}_{\AI,k}^{\mathsf H}\mathbf{f}_{\DD,j}|^2
&\le \newrev{b_k(q_k)}
\triangleq\epsilon_k^2q_k-\eta,\\
q_{k,\min}\triangleq\frac{\eta}{\epsilon_k^2}
&\le q_k\le P_{\mathrm{LO,th}}.
\end{aligned}
\label{eq:budget}
\end{equation}
At $q_k=q_{k,\min}$, the data-flow interference budget is zero, so each data beamformer must be orthogonal to client $k$. Increasing $q_k$ permits more data-flow interference at the AI client, but consumes BS power and can increase AI-stream interference at data users. We search over the finite set $\mathcal{V}=\{v_1,\ldots,v_{N_v}\}\subset[0,1]$ using the grid
\begin{equation}
q_k(v)=q_{k,\min}+v(P_{\mathrm{LO,th}}-q_{k,\min}),
\quad v\in\mathcal{V}.
\label{eq:power_grid}
\end{equation}
Feasibility requires $q_{k,\min}\le P_{\mathrm{LO,th}}$ for each AI client $k$.

\subsection{Update of AI Beamformer}
For fixed $v$, define $\mathbf{H}_{\AI}=[\mathbf{h}_{\AI,1},\ldots,\mathbf{h}_{\AI,K_{\AI}}]\in\mathbb{C}^{N_{\mathrm t}\times K_{\AI}}$ and $\mathbf{d}(v)\in\mathbb{C}^{K_{\AI}}$, whose $k$-th element is $\sqrt{q_k(v)}e^{\mathrm{j}\phi_k}$. We set $\phi_k=0$ as a common phase reference because the computing and mixer constraints depend only on the received magnitude, while $\beta_k$ compensates the received phase. We formulate the AI beamformer design problem as follows:
\begin{align}
\underset{\mathbf{f}_{\AI}}{\operatorname{minimize}}\quad&
\mathbf{f}_{\AI}^{\mathsf H}\mathbf{Q}\mathbf{f}_{\AI}
\notag\\
\operatorname{subject\ to}\quad&
\mathbf{H}_{\AI}^{\mathsf H}\mathbf{f}_{\AI}=\mathbf{d}(v),
\label{eq:aibeam}
\end{align}
where $\mathbf{H}_{\DD}=[\mathbf{h}_{\DD,1},\ldots,\mathbf{h}_{\DD,K_{\DD}}]\in\mathbb{C}^{N_{\mathrm t}\times K_{\DD}}$ and
$\mathbf{Q}=\mathbf{I}+\lambda_{\DD}\mathbf{H}_{\DD}\mathbf{H}_{\DD}^{\mathsf H}$.
The coefficient $\lambda_{\DD}$ penalizes AI beamformer energy in the data-user channel subspace. The equality-constrained solution is
\begin{equation}
\mathbf{f}_{\AI}
=\mathbf{Q}^{-1}\mathbf{H}_{\AI}
(\mathbf{H}_{\AI}^{\mathsf H}\mathbf{Q}^{-1}\mathbf{H}_{\AI})^{-1}
\mathbf{d}(v).
\label{eq:aiclosed}
\end{equation}
Thus, each target received power \octrev{produces} one AI beamformer.

\subsection{Update of Data Beamformers}
For fixed $\mathbf{f}_{\AI}$ and $v$, only the data beamformers remain to be optimized. From \eqref{eq:tx}, data user $j$ observes $y_j=\mathbf{h}_{\DD,j}^{\mathsf H}\mathbf{x}_{\mathrm{BS}}+n_j$, where the symbols $\{s_i\}_{i\in\schD}$ and $s_w$ are mutually uncorrelated with unit power, and $n_j\sim\mathcal{CN}(0,\sigma_j^2)$. Since each user $j$ has one antenna, a complex equalizer $u_j\in\mathbb{C}$ forms $\hat s_j=u_j^*y_j$. For fixed $\mathbf{f}_{\AI}$, define $\Psi_j(\mathbf{F}_{\DD})\triangleq\mathbb{E}[|y_j|^2]$ and $\mu_j(u_j,\mathbf{F}_{\DD})\triangleq\mathbb{E}[|\hat s_j-s_j|^2]$. Their arguments are omitted below for clarity. Expanding these quantities gives
\begin{align}
\Psi_j={}&
\sum_{i\in\schD}|\mathbf{h}_{\DD,j}^{\mathsf H}\mathbf{f}_{\DD,i}|^2
+|\mathbf{h}_{\DD,j}^{\mathsf H}\mathbf{f}_{\AI}|^2+\sigma_j^2,
\notag\\
\mu_j={}&|u_j|^2\Psi_j
-2\operatorname{Re}\{u_j^*\mathbf{h}_{\DD,j}^{\mathsf H}\mathbf{f}_{\DD,j}\}+1.
\label{eq:mse}
\end{align}
For fixed $\mathbf{F}_{\DD}$, minimizing the scalar \octrev{mean-square error (}MSE\octrev{)} $\mu_j$ over $u_j$ gives the minimum \octrev{MSE} (MMSE) equalizer $u_j^{\mathrm{mmse}}$ and the corresponding minimum MSE $\mu_j^{\mathrm{mmse}}$:
\begin{equation}
u_j^{\mathrm{mmse}}
=\frac{\mathbf{h}_{\DD,j}^{\mathsf H}\mathbf{f}_{\DD,j}}{\Psi_j},
\qquad
\mu_j^{\mathrm{mmse}}=\frac{1}{1+\gamma_j}.
\label{eq:mmse}
\end{equation}
With an auxiliary MSE weight $\omega_j>0$, the right-hand side of the rate-MSE identity \cite{shi2011wmmse} is maximized by $u_j=u_j^{\mathrm{mmse}}$ and $\omega_j=(\mu_j^{\mathrm{mmse}})^{-1}$:
\begin{equation}
\log_2(1+\gamma_j)
=\frac{1}{\ln 2}
\underset{u_j,\,\omega_j>0}{\operatorname{maximize}}
\left(\ln\omega_j-\omega_j \mu_j+1\right)\mathpunct{.}
\label{eq:wmmse_identity}
\end{equation}
Applying \eqref{eq:wmmse_identity} to all users and summing over $j$ expresses the aggregate rate in \eqref{eq:sumrate} as a joint maximization over $\mathbf{F}_{\DD}$, $\{u_j\}$, and $\{\omega_j\}$. For fixed $\{u_j,\omega_j\}$, $\ln\omega_j+1$ is constant, so optimizing the data beamformers reduces to minimizing $\sum_{j\in\schD}\omega_j\mu_j$. This convex quadratic objective yields the following quadratically constrained quadratic program (QCQP):
\begin{align}
\underset{\mathbf{F}_{\DD}}{\operatorname{minimize}}\quad&
\sum_{j\in\schD}\omega_j \mu_j
\notag\\
\operatorname{subject\ to}\quad&
\sum_{j\in\schD}\|\mathbf{f}_{\DD,j}\|^2\le P_{\max}-\|\mathbf{f}_{\AI}\|^2,
\notag\\
&\sum_{j\in\schD}|\mathbf{h}_{\AI,k}^{\mathsf H}\mathbf{f}_{\DD,j}|^2
\le\newrev{b_k(q_k(v))},\quad k\in\setA\mathpunct{.}
\label{eq:wmmse}
\end{align}
Expanding the weighted sum-MSE objective in \eqref{eq:wmmse} and collecting the terms that depend on the data beamformers gives $\mathbf{B}=\sum_{i\in\schD}\omega_i|u_i|^2\mathbf{h}_{\DD,i}\mathbf{h}_{\DD,i}^{\mathsf H}\in\mathbb{C}^{N_{\mathrm t}\times N_{\mathrm t}}$ and $\mathbf{c}_j=\omega_j u_j\mathbf{h}_{\DD,j}\in\mathbb{C}^{N_{\mathrm t}}$. Let $\lambda\ge0$ denote the Lagrange multiplier for the total-power constraint, and let $\nu_k\ge0$ denote the Lagrange multiplier for the data-flow interference constraint of AI client $k$. Define $\boldsymbol{\nu}=[\nu_1,\ldots,\nu_{K_{\AI}}]^{\mathsf T}\in\mathbb{R}_{+}^{K_{\AI}}$. The Karush-Kuhn-Tucker (KKT) stationarity conditions give
\begin{subequations}\label{eq:kkt}
\begin{align}
\mathbf{f}_{\DD,j}&=\mathbf{A}^{-1}(\lambda,\boldsymbol{\nu})\mathbf{c}_j,\quad j\in\schD,
\label{eq:kkt_beam}\\
\mathbf{A}(\lambda,\boldsymbol{\nu})
&=\mathbf{B}+\lambda\mathbf{I}
+\sum_{k\in\setA}\nu_k\mathbf{h}_{\AI,k}
\mathbf{h}_{\AI,k}^{\mathsf H}.
\label{eq:kkt_matrix}
\end{align}
\end{subequations}
For fixed $\boldsymbol{\nu}$, the total beamforming power in \eqref{eq:kkt} is non-increasing in $\lambda$. We set $\lambda=0$ when the power constraint is inactive and otherwise find the power-feasible value by bisection. Each $\nu_k$ is updated by a projected dual step proportional to its normalized data-flow interference violation. A violation therefore increases the spatial penalty along that client's channel. The common Hermitian matrix $\mathbf{A}(\lambda,\boldsymbol{\nu})$ in $\eqref{eq:kkt_matrix}$ is factored once for all $K_{\DD}$ data beamformers. Alternating projections remove small interference and power residuals.

\subsection{Computational Complexity}
Let $N_v=|\mathcal V|$, and let $I_{\mathrm W}$ and $I_{\mathrm d}$ denote the numbers of WMMSE iterations and dual updates, respectively. For each received-power candidate, factorizing the common $N_{\mathrm t}\times N_{\mathrm t}$ matrix in \eqref{eq:kkt} costs $\mathcal{O}(N_{\mathrm t}^{3})$. The $K_{\DD}$ back substitutions and $K_{\AI}$ constraint evaluations contribute $\mathcal{O}(N_{\mathrm t}^{2}K_{\DD}+N_{\mathrm t}K_{\DD} K_{\AI})$. The leading complexity is thus $\mathcal{O}(N_vI_{\mathrm W}I_{\mathrm d}[N_{\mathrm t}^{3}+N_{\mathrm t}^{2}K_{\DD}+N_{\mathrm t}K_{\DD} K_{\AI}])$. The factorization is reused across data beamformers, and the $N_v$ candidates can be evaluated in parallel. 

\begin{figure*}[!t]
\centering
\begin{minipage}[t]{\resultpanelwidth}
\centering
\includegraphics[hiresbb,width=0.92\linewidth]{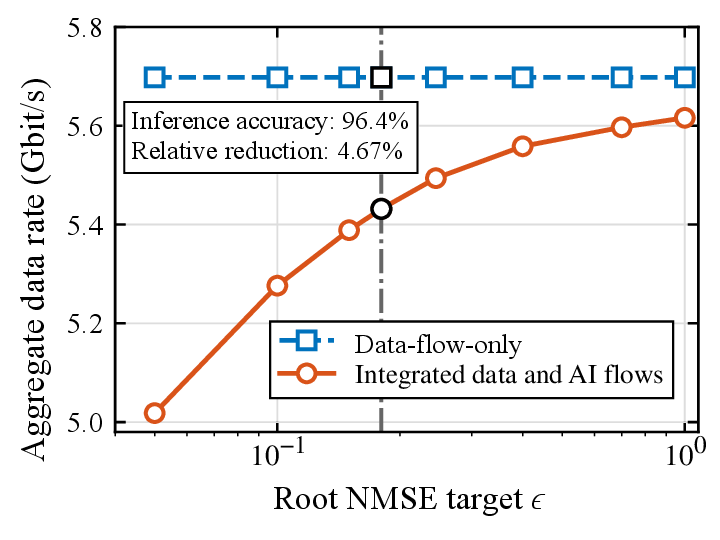}
\vspace{-12pt}
\caption{Aggregate data rate versus $\epsilon$.}
\label{fig:epsilon}
\end{minipage}\hspace{0.013\textwidth}%
\begin{minipage}[t]{\resultpanelwidth}
\centering
\includegraphics[hiresbb,width=0.92\linewidth]{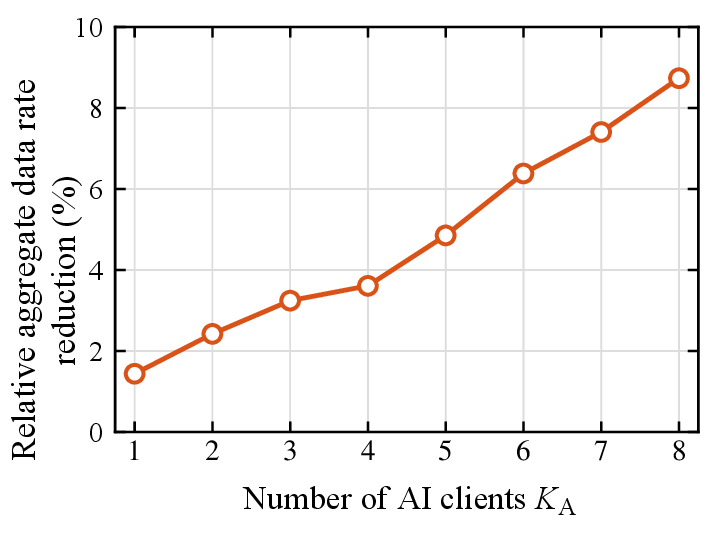}
\vspace{-12pt}
\caption{Relative aggregate data rate reduction versus $K_{\AI}$.}
\label{fig:kai}
\end{minipage}\hspace{0.014\textwidth}%
\begin{minipage}[t]{\resultpanelwidth}
\centering
\includegraphics[hiresbb,width=0.92\linewidth]{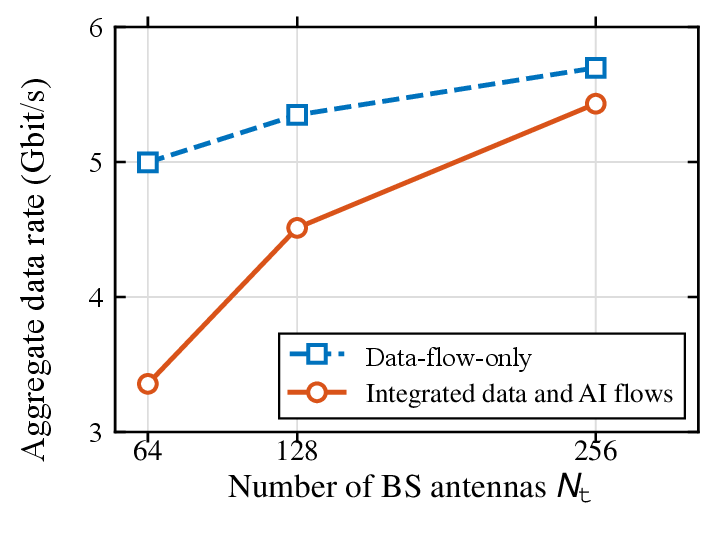}
\vspace{-12pt}
\caption{Aggregate data rate versus $N_{\mathrm t}$.}
\label{fig:nt}
\end{minipage}\hspace{0.013\textwidth}%
\begin{minipage}[t]{\resultpanelwidth}
\centering
\includegraphics[hiresbb,width=0.92\linewidth]{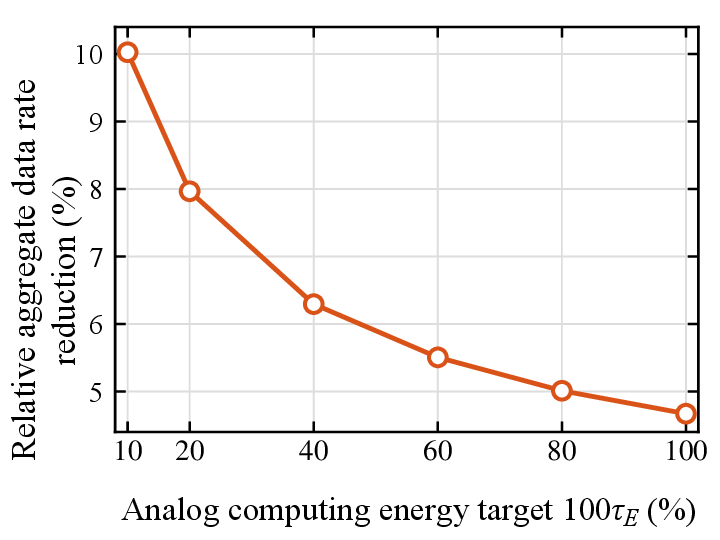}
\vspace{-12pt}
\caption{Relative aggregate data rate reduction versus $\tau_E$.}
\label{fig:energy}
\end{minipage}
\par\vspace{4pt}
\end{figure*}

\begin{table}[t]
\caption{Analog RF Computing Parameters}
\label{tab:params}
\centering
\scriptsize
\renewcommand{\arraystretch}{0.86}
\setlength{\tabcolsep}{2.2pt}
\begin{tabular}{@{}p{0.18\columnwidth}p{0.49\columnwidth}p{0.25\columnwidth}@{}}
\toprule
Parameter & Description & Value \\
\midrule
$N_{\mathrm{FFT}}$ & Fast Fourier transform (FFT) size & $4096$ \\
$B$ & Occupied bandwidth & $25$ MHz \\
$T_0$ & Environmental temperature & $300$ K \\
$\vartheta$ & Guard factor & $0.33$ \\
$M'$ & Size of row blocks & $6$ \\
$\varpi$ & CP overhead & $0.125$ \\
$P_{w,0}$ & BS-side reference power level & $0$ dBm \\
$P_{x,0}$ & Client-side reference power level & $0$ dBm \\
$P_{x,\max}$ & Client transmit power limit & $23$ dBm \\
$P_{\mathrm{LO,th}}$ & LO-port saturation threshold & $-3$ dBm \\
$\rho_{\mathrm{radio}}$ & Client RF-chain efficiency & $0.3$ \\
$\rho_{\mathrm{mixer}}$ & Mixer conversion coefficient & $0.2512$ \\
$\rho_{\mathrm{nf}}$ & Effective receiver noise coefficient & $0.2512$ \\
$e_{\mathrm{adc}}$ & ADC energy per sample & $1$ pJ/sample \\
$e_{\mathrm{dec}}$ & Digital energy per readout operation & $1$ pJ/operation \\
$e_{\mathrm{digital}}$ & Digital computing energy per MAC & $3$ pJ/MAC \\
\bottomrule
\end{tabular}
\end{table}

\section{Performance Evaluation}
For performance evaluation, we consider the 3GPP Technical Report (TR) 38.901 channel model for an indoor factory with sparse clutter and a high base station (InF-SH) \cite{3gpp38901}. For carrier frequency $f_c$ in GHz and three-dimensional (3-D) distance $d_{\mathrm{3D}}$ in meters, the line-of-sight (LOS) and non-line-of-sight (NLOS) path losses are
\begin{align}
\mathrm{PL}_{\mathrm{LOS}}
&=31.84+21.5\log_{10}d_{\mathrm{3D}}+19\log_{10}f_c,
\notag\\
\mathrm{PL}_{\mathrm{NLOS}}
&=\max\{\mathrm{PL}_{\mathrm{LOS}},
32.4+23\log_{10}d_{\mathrm{3D}}+20\log_{10}f_c\}.
\label{eq:inf_pathloss}
\end{align}
The InF-SH LOS probability is $p_{\mathrm{LOS}}(d_{\mathrm{2D}})=\exp(-d_{\mathrm{2D}}/k_{\mathrm{SH}})$ with $k_{\mathrm{SH}}=582.6$ m. LOS and NLOS shadowing deviations are $4$ and $5.9$ dB, respectively. LOS links use a $9$ dB Rician channel over a uniform planar array (UPA), while NLOS links are Rayleigh. The BS and user heights are $8$ m and $1.5$ m, respectively. The horizontal distance $d_{\mathrm{2D}}$ is uniformly distributed over $[10,15]$ m, and the 3D distance used for path loss is $d_{\mathrm{3D}}=\sqrt{d_{\mathrm{2D}}^2+(6.5\,\mathrm{m})^2}$, where $6.5$ m is the BS-user height difference.  The BS and user antenna gains are set as $8$ dBi and $3$ dBi, respectively. The fully digital BS uses a half-wavelength $16 \times 16$ UPA with $N_{\mathrm t}=256$ to serve $K_{\DD}=7$ data users, and transmit one AI stream to $K_{\AI}=4$ AI clients. The shared power is set as $P_{\max}=0.25N_{\mathrm t}$ W. The integrated-data-and-AI-flows case and the data-flow-only case use identical data users and channels in each sweep. 

The analog RF computing model follows \cite{yu2026analogrf}. We use the second LeNet convolutional layer \cite{lecun1998gradient}, with $M=16$ and $N=150$ as a representative NN layer. The other key parameters are listed in Table~\ref{tab:params}. Equations~\eqref{eq:energy_input}--\eqref{eq:energy_read} give $\xi=4.9875\times10^{-8}$ s, $e_2=4.43$ fJ/MAC, and $e_3=13.28$ fJ/MAC. Hence, we have $e_{\AI,k}=49.875|\beta_k|^2+0.01772 \text{pJ/MAC}$. The same values give $\kappa=6.09\times10^5$. The RF port limit gives $\beta_{\max}=14.13$, while the energy constraint gives $\bar\beta=0.2445$ at $\tau_E=1$ and $0.0752$ at $\tau_E=0.1$. We use $\lambda_{\DD}=10^6$ and $\mathcal V=\{0,0.005,0.02,0.08,0.2,0.5,0.9\}$.

The task-level calibration in \cite{yu2026analogrf} injects layer-output errors and evaluates LeNet on the Modified National Institute of Standards and Technology (MNIST) dataset. Unless otherwise stated, we set $\epsilon=0.18$, which is the largest calibrated error tolerance that maintains at least $96\%$ accuracy. 

\subsection{Impact of Computing Accuracy Target}
Fig.~\ref{fig:epsilon} plots the aggregate data rate as a function of the computing accuracy target. At $\epsilon=0.18$, the relative aggregate data rate reduction is $4.67\%$\octrev{. At the leftmost plotted point,} $\octrev{\epsilon}\mathrel{\octrev{=}}0.05$\octrev{, the integrated and data-flow-only rates are $5.02$ and $5.70$ Gbit/s, respectively, giving an} $11.93\%$ reduction. From \eqref{eq:budget}, the minimum received computing power scales as $1/\epsilon^2$. Its value at the marked point is only $7.7\%$ of that at $\epsilon=0.05$. Once the required inference accuracy is met, a tighter target consumes excessive power but only offers negligible improvements in the accuracy of AI inference. The computing accuracy target of analog MVMs should be set as the largest error tolerance that can satisfy the required inference accuracy. 

\subsection{Impact of Number of AI Clients}
Fig.~\ref{fig:kai} plots the relative aggregate data rate reduction as a function of the number of AI clients. We cover one to eight clients, over which the relative aggregate data rate reduction increases from $1.44\%$ to $8.74\%$. Although all AI clients share one multicast AI stream, each additional client introduces a computing-gain constraint and an interference constraint. The nonuniform increments reflect the channel alignment. Grouping clients with spatially compatible channels can therefore lower the aggregate data rate reduction caused by integrating the AI flow with existing data flows.

\subsection{Impact of Spatial Degrees of Freedom}
Fig.~\ref{fig:nt} compares the data-flow-only case with integrated data and AI flows as the number of antennas $N_{\mathrm t}$ varies, with $K_{\DD}=7$. The relative aggregate data rate reduction decreases from $32.83\%$ at $N_{\mathrm t}=64$ to $4.67\%$ at $N_{\mathrm t}=256$. The two lines approach each other as $N_{\mathrm t}$ increases.  These results indicate that additional spatial degrees of freedom can allow the data beamformers to suppress their interference at AI clients and also let the AI beamformer avoid interfering with the data users at the same time. 

\subsection{Impact of Client Energy Target}
Fig.~\ref{fig:energy} shows the effect of $\tau_E$, the analog computing energy target relative to the energy consumption of AI inference on digital computing hardwares, i.e., GPUs. Reducing $\tau_E$ from $1$ to $0.1$ lowers $\bar\beta$ and raises the relative aggregate data rate reduction from $4.67\%$ to $10.03\%$. A smaller client gain shifts the computing burden to the BS and exposes a direct tradeoff between client energy and aggregate data rate. The most significant raise of aggregate data rate happens between $10\%$ and $40\%$, after which the data rate enhancement diminish. The client energy target should be carefully chosen for a good tradeoff between communication and computing. 

\section{Conclusion}
We studied integrated data and AI flows in which a BS integrates an analog RF computing-based AI flow to an existing wireless  communication system with multiple data flows. We developed joint beamforming under computing accuracy, client energy, BS power, and mixer requirements. The results show that the relative aggregate data rate reduction depends on the inference target, available spatial degrees of freedom, number of AI clients, and client energy target. Numerical results indicate that telecom operators can multicast an AI model to multiple edge devices for on-device inference while largely preserving data communication performance. Future work will consider multiple AI flows (i.e., serving multiple AI models at the BS), client scheduling, \octrev{moving clients such as those in low-altitude wireless networks \cite{zhang2026multimodal},} and hardware validation of the system.

\let\originalbibitem\bibitem
\def\newreferencekey{zhang2026multimodal}
\renewcommand{\bibitem}[1]{\originalbibitem{#1}\def\currentreferencekey{#1}\ifx\currentreferencekey\newreferencekey\color{currentrevisioncolor}\else\color{black}\fi}
\bibliographystyle{IEEEtran}
\bibliography{references}

\end{document}